\documentclass{aa}
\usepackage{graphicx}
\usepackage{natbib}
\usepackage{color}
\bibpunct{(}{)}{;}{a}{}{,} 
\begin{document}

\def\ms{M$_{\odot}$}
\def\zs{Z$_{\odot}$}
\def\mi{M$_{\rm IN}$}
\def\zi{Z}
\def\xi{Z$_{\rm IN}$}
\def\me{M$_{\rm ENV}$}
\def\nsn{N$_{\rm Ib,c}$/N$_{\rm II}$}
\def\mup{M$_{\rm Ib,c}$}
\def\lb{L$_{\rm B,\odot}$}
\def\be{$\beta^+$}
\def\ee{e$^+$}
\def\ne{N$_{e^+}$}
\def\co{$^{56}$Co }
\def\ti{$^{44}$Ti }
\title{On the intensity and spatial morphology of the
511 keV emission in the Milky Way}

\author{ Nikos Prantzos  }

\authorrunning{N. Prantzos}
 
\titlerunning{On the Galactic 511 keV emission}

\offprints{N. Prantzos}

\institute{ Institut d'Astrophysique de Paris, 98bis Bd. Arago, 
              75104 Paris, France,  
              \email{prantzos@iap.fr}
           }
\date{accepted : December 10, 2005}

\abstract{The positron emissivity of the
Galactic bulge and disk, resulting from radioactivity of SNIa, is
reassessed in the light of a recent evaluation of the SNIa rate. 
It is found that the disk may supply more positrons than
required by recent SPI/INTEGRAL observations, but the bulge
(where the characteristic \ee \ annihilation line at 511 keV  
is in fact observed) only about 10\%.
It is argued  that a large fraction of the disk positrons
may be transported via the regular magnetic field of the Galaxy into the bulge,
where they annihilate. This would increase both the bulge positron
emissivity and the bulge/disk ratio, alleviating considerably 
the constraints imposed by INTEGRAL data analysis. We argue that the bulge/disk
ratio can be considerably smaller than the values derived by the recent 
analysis of Kn\"odlseder et al. (2005), if the disk positrons diffuse 
sufficiently away from their sources, as required by our model; 
this possibility could be soon tested, as data are accumulated in 
the SPI detectors. The success of the proposed  scenario depends 
critically upon
the,  very poorly known at present, properties of the galactic magnetic field
and of the propagation of low energy positrons in it.
}
\maketitle

%

\section{Introduction}

The origin of the Galactic electron-positron annihilation radiation
has remained problematic ever since the original detection of
its characteristic 511 keV line (e. g. Diehl et al. 2005 and references
therein). In particular, recent  observations  
of the line intensity and spatial morphology with the SPI instrument aboard
INTEGRAL put severe constraints on its origin, since
it appears that 1.5$\pm$0.1 10$^{43}$ \ee/s are annihilated in the bulge alone
and  0.3$\pm$0.2 10$^{43}$ \ee/s  in the disk, i.e. that
the bulge/disk positron emissivity  ratio is $B/D\sim$5 
(Kn\"odlseder et al. 2005). The most promising sources of 
positrons in the Galaxy apear to be type Ia supernovae, but 
``conventional'' models of Galactic \ee \
production through SNIa radioactivity  fail to explain these features
(see Sec. 2). This failure prompted suggestions of more ``exotic''
models, involving SNIc supernovae and 
$\gamma$-ray bursts (Nomoto et al. 2001, Cass\'e et al. 2004, 
Bertone et al. 2004, Parizot et al. 2004), low mass X-ray binaries 
(Prantzos 2004), millisecond pulsars (Wang et al. 2005), microquasars 
(Guessoum et al. 2005), and more exotic ones, like
annihilation of light dark matter particles
(Boehm et al. 2004, Ascasibar et al. 2005) and a tangle of light
superconductiong cosmic strings (Ferrer and Vachaspati 2005); 
see Prantzos (2004), Diehl et al. (2005) or
Kn\"odlseder et al. (2005) for a critical
discussion of those suggestions.

In this work we reasses the SNIa \ee \ emissivity of the
Galactic bulge and disk in the light of recent data, and we find
that the sum of the two components is slightly larger than required from 
SPI measurements.  This may be a coincidence, but we
argue here that a  large fraction of the disk positrons
may be transported via the regular magnetic field of the Galaxy into the bulge,
where they annihilate. This increases both the bulge positron 
emissivity and the
bulge/disk ratio, alleviating considerably  the constraints imposed by 
the recent SPI/INTEGRAL data analysis (Kn\"odlseder et al. 2005). In fact, 
we argue that the SPI data are compatible with values of $B/D$ as low as 0.5, 
because positrons can propagate away from their sources and fill a rather 
large volume, much larger than the relatively thin disks adopted in 
the analysis of  Kn\"odlseder 
et al. (2005). This property is crucial to the success 
of the scenario proposed here, which depends also on the poorly known
 properties of the Galactic  magnetic field. 

In Sec. 2, the rate of positron production from SNIa in the Milky Way 
bulge and 
disk is evaluated and compared to the observations. In Sec. 3, the 
morphology of
 the Galactic magnetic field is discussed (Sec. 3.1), as well as 
several aspects
 of the propagation of positrons in it (Sec. 3.2, 3.3 and 3.4); in particular, 
it is argued that positrons can escape from the disk into the bulge and 
annihilate there. The spatial morphology of the 511 keV emission resulting 
from  such a transfer may, under certain conditions, be fully compatible 
with current observations.  In Sec. 4 we show that the bulge/disk 
positron emissivity ratio
 may be as low as 0.5 and still compatible with SPI data. We substantiate 
that claim by  calculating the resulting flux morphologies and intensities 
for several plausible distributions of the disk positrons, assumed to diffuse
 in a large volume (akin to the "Cosmic Ray Halo", occupied by ordinary 
cosmic rays). The existence of such a low surface brightness 
"Cosmic positron halo" 
could be put to test in a few years, as more data are accumulated in the SPI
 detectors; if it is confirmed, it will have important implications for our 
understanding of positron production, propagation and annihilation in 
the Galaxy 
(independently of the model of positron transfer from the disk to the
 bulge proposed here).

\section{Positron production from SNIa in the Galaxy}

A comprehensive overview of ``conventional''
positron production  sites in the Galaxy
has been presented in several places (Dermer and Murphy 2000, Prantzos 2004,
Diehl et al. 2005, Kn\"odlseder et al. 2005). The most prominent source
of galactic positrons 
 appears to be the \be \ radioactivity of supernovae (SN)
and, in particular, beta decay  of \co produced in thermonuclear SN (SNIa). 
The peak luminosity of those objects suggests that they
produce, on average, $\sim$0.7 \ms \ of \co ($\sim$1.5 10$^{55}$ nuclei,
releasing \ee \ in $\sim$20\% of their decays).
However, because of the relatively short
lifetime of \co and of the poorly understood configuration of 
supernovae magnetic fields, theoretical estimates of
the positron escape fraction $f_{56}$ are
extremely uncertain at present (e.g. Chan and Lingenfelter 1993).
Observations offer, in principle,  
a much more reliable way to evaluate  $f_{56}$, through
the shape of the late optical lightcurve of SNIa. A pioneering study
(Milne et al. 1999) of a sample of  SNIa concludes that 
\ne=8$^{+8}_{-4}$ 10$^{52}$ positrons
escape from  an average SNIa, i.e. that $f_{56}\sim$3\%.
In the following, we shall adopt this as a canonical value, although
a recent study of  the thermonuclear SN2000cx (Sollerman et al. 2004), covering
optical and near-IR wavelengths, concludes that its late lightcurve
is compatible with $f_{56}\sim$0.   
However, as Sollerman et al. (2004) recognize, ``...these conclusions
are drawn from observations of a single SN, which was clearly
unusual at the peak... and they have to be verified by more data..''.


The next important ingredient in order to evaluate the \ee 
production rate in a galactic system is the SNIa rate $R_{Ia}$. 
Most previous
studies evaluated that rate in terms of SN frequency per unit
B-band luminosity (e.g. Cappellaro 2003), which is a 
a poor tracer of the stellar mass of a system.
Mannucci et al. (2004) use the complete catalogue of 
near-IR galaxy magnitudes obtained by the 2MASS survey
to evalute SN frequencies
per unit  luminosity in th near IR
(which is a much better tracer of stellar mass),
as a fuction of galaxian morphological type.
They find that, in units of
[100 yr 10$^{10}$ \ms]$^{-1}$ (SNuM) $R_{Ia}$
is :  0.044$^{+0.016}_{-0.014}$ for E/S0, 0.065$^{+0.027}_{-0.025}$
for Sa/b and 0.17$^{+0.068}_{-0.063}$  for Sbc/d, 
i.e. a factor $\sim$4 higher in 
late spirals than in ellipticals.  Note that those values are 
systematically higher (a factor $\sim$2) than previous estimates
(e.g. in Cappellaro et al. 2003).

The last ingredient is the mass of the galactic system. In the case of
the Milky Way bulge, various studies converge to values in the range
1-2 10$^{10}$ \ms, either through photometric (e.g. Robin et al. 2004,
Dwek et al. 1995) or dynamical (e.g. Klypin et al. 2003) determinations. 
We adopt
M$_B$=1.5$^{+0.5}_{-0.5}$ 10$^{10}$ \ms \ in the following. 
Similar uncertainties exist in the case of the
Milky Way disk (e.g. Boissier and Prantzos
1999, Robin et al. 2004).
We adopt M$_D$=4.5$^{+1.5}_{-1.5}$ 10$^{10}$ \ms \ in the following.

The \ee \ production rate from \co radioactivity of SNIa is then:
$$
S \ = \ M \ R_{Ia} \ N_{e^+} 
$$
In the case of the Galactic disk, clearly identified as a Sb/c spiral,
one obtains  $S_D$=1.95$^{+0.98}_{-0.93}$ 10$^{43}$
\ee/s. \footnote{Errors are assumed to have gaussian 
distributions and add quadratically, but only $M$ and  $R_{Ia}$ are used
in the error       calculation. Uncertainties of $N_{e^+}$ are rather 
systematic and have not been taken into account. One may include them 
formally as e.g.  $S_D$=1.95$^{+0.98}_{-0.93}$ $^{+1.95}_{-0.98}$  10$^{43}$
\ee/s.}
The Galactic bulge is usually assumed to have the morphology of an early type
galaxy (E/S0) and in that case one obtains  
$S_B$=0.17$^{+0.083}_{-0.081}$ 10$^{43}$
\ee/s. Those conservative estimates suggest that:
1) the bulge
\ee \ emissivity is between 0.06 and 0.16 of the one inferred from 
SPI mesurements; 2) the disk has $\sim$12 time more
SNIa than the bulge, and correspondingly larger \ee \ emissivity; 3) the disk
emissivity is  slightly larger than  the total galactic (mostly bulge)
emissivity required by SPI observations.

Thus, {\it assuming that the adopted $N_{e^+}$ is correct}, one sees that
SNIa in the Galaxy may indeed provide the positron emissivity required
by observations; however, while theory suggests in that case
a large Disk/Bulge ratio
$D/B\sim$10, observations show exactly  the opposite: $D/B\sim$0.2.
This a standard problem encountered by almost any one of the suggested
positron sources, with the exception of the dark matter and of the tangle
of light superconducting strings; they cannot, in general,  
reproduce the morphology of the 511 keV emission observed by SPI, although some
of them (e.g. X-ray binaries, microquasars) may encounter less 
difficulty than others.
Note, however, that
for most of those sources, their Galactic rates and positron emissivities are
much more uncertain than for SNIa.

In fact, the problem may be even worse. As noted in Kn\"odlseder et al. (2005),
the disk emissivity may be entirely explained by the positrons
released from the decay of $^{26}$Al, a radioactive nucleus produced
in massive stars, with a half-life of $\sim$1 Myr. Its characteristic
gamma-ray line at 1.8 MeV has detected in the plane of the Milky Way
by various instruments in the past 20 years (see Prantzos 1991, or
Prantzos and Diehl 1996 and references therein), with a flux corresponding
to the decay of $\sim$3 \ms \ of  $^{26}$Al per Myr.
The corresponding positron emissivity could explain most (if not all)
of the disk 511 keV flux, at least in the thin disk models tested by 
Kn\"odlseder et al. (2005). However, as we argue in Sec. 4, the $^{26}$Al
positron emissivity may well be accomodated in the framework of the
scenario proposed here, which involves an extended disk of positrons
from other sources, like SNIa.

If the Galactic SNIa positron emissivity evaluated in this section
is close to the 
real onen but the source of the observed bulge emission turns out to
be  different, it should then  be a rather strange coincidence.
We argue below that transport of disk positrons to the bulge through the
Galactic magnetic field may inverse the Disk/Bulge ratio.
The arguments  are valid for any other source producing positrons of 
$\sim$1 MeV, such as those resulting from radioactivity.

\section{Positron propagation/annihilation in the Galaxy}

In the Milky Way disk, 
the scaleheight of the most prolific \ee sources,
namely SNIa, should be similar to the one of the old stellar population of the
disk. Recent studies (e.g. Chen et al. 2001, Siebert et al. 2004)
find $H_*\sim$330-350 pc in the solar neighborhood, while in the inner Galaxy
it is smaller but never drops below 300 pc (e.g. Ferri\`ere 1998, Narayan and 
Jog 2002 and references therein). On the other hand, positrons slow down and
annihilate in the gaseous medium of the disk, which has a smaller scaleheight
($\sim$100 pc in the solar neighborhood,
Ferri\`ere 1998). Fig. 1 summarises the relevant data (the profile of the SNIa
rate is from the Milky Way evolutionary model of Boissier and Prantzos 1999).

\begin{figure}
\centering
\includegraphics[angle=-90,width=0.5\textwidth]{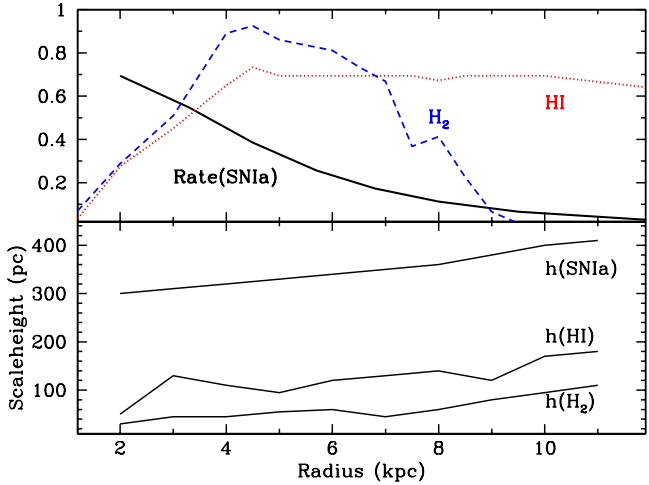}
\caption{{\it Top}:
Radial surface densities of H$_2$ ({\it dashed} curve), HI ({\it dotted})
and SNIa rate  ({\it solid})
in the Milky Way; the first two are expressed in \ms/pc$^2$ and 
in logarithmic scale,
while the latter is in (Gyr pc$^2$)$^{-1}$ and in linear scale. 
{\it Bottom}: Scale heights of H$_2$, HI and SNIa
as a function of galactocentric radius (see text for references).
} 
\end{figure}

\begin{figure}
\centering
\includegraphics[angle=-90,width=0.5\textwidth]{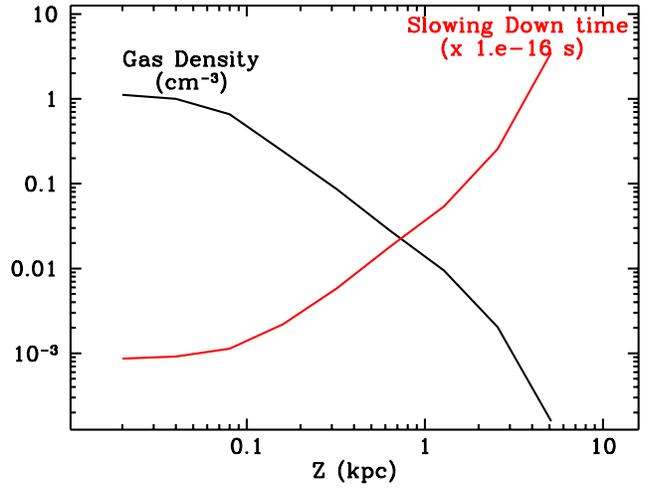}
\caption{\label{f2} Total gas density and timescale for 
positron slow down 
as a function of distance from the Galactic plane at galactocentric distance
8 kpc; curves for other galactocentric distances differ little from those
ones. The timescale is calculated from the corresponding
timescales for the various gaseous phases, i.e. $\tau_{SD}^{-1} \ = \ \Sigma 
(f_i\tau_i)^{-1}$, where $i$ stands for  cold neutral, warm neutral,
warm ionised and hot ionised gas, and the corresponding density profiles
and volume fiiling factors $f_i$ are taken from Ferri\`ere (1998).
}  
\end{figure}

Thus, a large fraction of the positrons released by the SNIa of the Galactic 
disk are found in a medium of substantially lower density $n$ 
than the  local Galactic  plane. 
Radioactivity positrons have energies $\sim$1 MeV and 
lose energy mostly through Coulomb  interactions, with a
characteristic timescale of
$\tau_{SD}\sim$10$^5$ $(n/{\rm cm^3})^{-1}$ yr (e.g. Forman, Ramaty and Zweibel 1986).
For  typical gas densities outside the neutral gas layer of 
the Milky Way disk, $\tau_{SD}$ is  larger than 10$^6$ yr, as can be seen in 
Fig. 2 (calculated taking into account the 
various gaseous phases and corresponding
volume filling factors of the ISM from Ferri\`ere 1998).
Thus, it appears that positrons of $\sim$MeV energies resulting 
from SNIa radioactivity in the Milky Way disk 
wander mostly through a low density ISM for a couple of Myr, before 
thermalisation and annihilation (see also Sec. 3.2).

\begin{figure}
\centering
\includegraphics[width=0.5\textwidth]{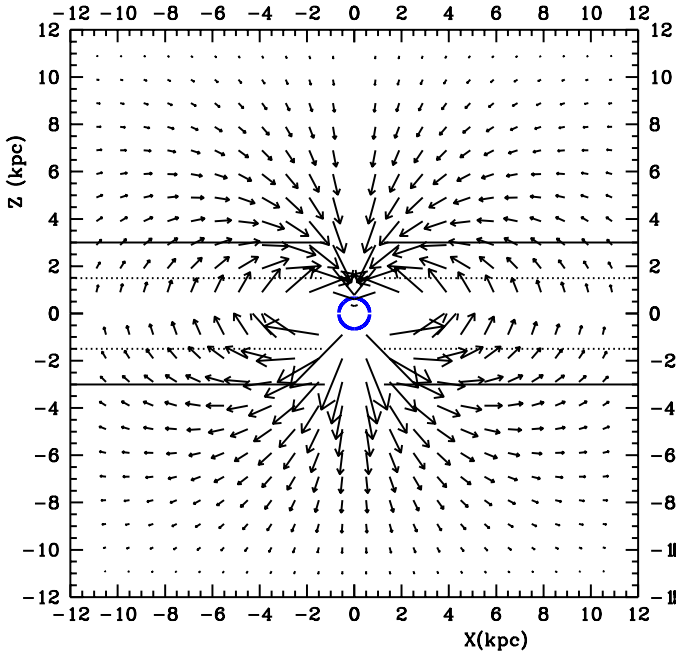}
\caption{\label{f3} A representation of the poloidal magnetic field of
the Milky Way in the x-z plane, {\it assuming} it is a A0 dipole. 
The central circle corresponds to the size of the 511 keV emitting
region (FWHM) seen by SPI aboard INTEGRAL. The irregular (turbulent)
magnetic field {\it is assumed} negligible outside the 
region between the horizontal lines ({\it dotted} 
lines are suggested in Prouza and Smida 2003, while {\it solid} lines are
lower limits to the cosmic ray halo size suggested in
Strong and Moskalenko 2001); positrons escaping from that region are 
presumably directed
by the dipole field lines towards the bulge.
}  
\end{figure}

\subsection{The Galactic magnetic field}

During those long timescales, positrons move through the magnetic field (MF) 
of the Milky Way. The configuration of the Galactic MF is 
probed mainly through measurements of Faraday rotation of the radiation
emitted by pulsars and extragalactic radio sources (e.g. Vall\'ee 2004)
and it is poorly known today. It appears though that the large
scale regular MF is composed of a toroidal (disk) component (probably
bisymmetric) and a poloidal (halo) component, probably in the form
of a A0 dipole (see Han 2004 and references therein). For the latter
component we adopt (Fig. 3) recent
parametrisations (Alvarez-Muniz, Engel and Stanev, 2002; 
Prouza and Smida, 2003)
expressing its cartesian ($X,Y,Z$) components in cylindrical coordinates 
($r,\theta,\phi$):
$$
B_X \ = \ -3 \mu_G \ sin\theta \ cos\theta \ cos\phi \ / \ r^3 
$$
$$
B_Y \ = \ -3 \mu_G \ sin\theta \ cos\theta \ sin\phi \ / \ r^3 
$$
$$
B_Z   \  = \ \ \  \ \ \mu_G \ (1 \ - \ 3 \ sin^2\theta) \ / r^3 
$$
where $\mu_G$=184 $\mu G$ kpc$^3$ is the 
magnetic moment of the Galactic dipole. 
At Galactocentric distance r=8 kpc the toroidal component
has a strength of a few $\mu$G (Beck et al. 1996) and dominates
the poloidal one (a few tenths of  $\mu$G). However, the former varies 
as 1/r, while the latter as 1/r$^3$ and should therefore
dominate in the inner Galaxy.

Positron propagation is strongly affected by the
irregular (turbulent) component of the  galactic MF, which is comparable
in intensity with the regular one near the local disk. Unfortunately, the
properties of the irregular component away from the disk plane are even 
less well understood than those of the regular components (e.g. Han 2004).
In a recent work, Prouza and Smida (2003) assume that the 
turbulent component occupies 80\% of the volume inside spiral arms,
20\% of the volume outside spiral arms and within
vertical distance $|z|<$1.5 kpc and 
only 1\% of the volume at larger distances. 
On the other hand, cosmic ray propagation models
indicate that the size of the cosmic ray ``halo'' (CRH, i.e. the
region inside which cosmic rays diffuse on inhomogeneities of
the magnetic field) is $z_{CRH}>3$ kpc, based on
measurements of unstable/stable ratios of secondary nuclei
(Strong and Moscalenko 2001). If \ee \ escape from the CRH
then positron propagation at large distances from the disk will be dominated
by the regular MF, i.e. the poloidal field. In those conditions, a
fraction of the positrons  produced from disk SNIa will
ultimately find their way into the bulge.

For the model to work, the positrons have to avoid the ``mirror effect''
(i.e. bouncing back in the gradient of the poloidal magnetic field as they 
approach the bulge), while conditions in the bulge much be such that
positrons indeed annihilate there and do not escape it. Those questions
are discussed in Sec. 3.3 and 3.4, respectively.

As mentioned in Sec. 1 the SPI data analysis by Kn\"odlseder et al. (2005 and
private communication) slightly favour a composite model, with a bulge positron
luminosity of $\sim$1.5  10$^{43}$ \ee/s and a disk luminosity
of 0.3-0.5 10$^{43}$ \ee/s; in fact, as we discuss in Sec. 4, 
bulge and disk luminosities of 1.2 10$^{43}$ \ee/s each
may be compatible with the data, provided the disk is
sufficiently extended. 
In order to explain the observed bulge emissivity
by the proposed model, the fraction of positrons channeled to the bulge
must then be $f_{ESC}\sim$0.5 (taking into 
account the disk and bulge emissivities evaluated in Sec. 2), i.e.
$\sim$10$^{43}$ \ee/s from the disk have to join the 0.17 10$^{43}$ \ee/s 
produced in the bulge and annihilate in that region.

\subsection{Positron escape vs. confinement}

A ``naive'', first order estimate of the escaping fraction
of positrons $f_{ESC}$ may be obtained by
\begin{equation}
\frac{f_{ESC}}{f_{CONF}} \ \sim \ \frac{\tau_{SD}}{\tau_{CONF}}
\end{equation}
where $f_{CONF}$ is the fraction of positrons confined in the disk
($f_{CONF}$+$f_{ESC}$=1), $\tau_{SD}$ is the slow-down time for positrons
and $\tau_{CONF}$ is their confinement time in the disk, in the framework
of the Leaky Box model for cosmic ray propagation. 
Typical values are:  for the positron slow down timescale at
z=350 pc,  $\tau_{SD}\sim$1.5 10$^6$ yr (Fig. 2); and 
for the confinement time in the
CRH $\tau_{CONF}$=1.45$\pm$0.15 10$^7$ yr, 
a value obtained for normal cosmic rays 
(e.g.  Mewaldt et al. 2001). Thus, a first estimate
suggests that  only $\sim$10\%
of the disk positrons may escape before slowing down.

We feel, however,  that the many uncertainties of the problem require a 
much more detailed investigation and that a larger escape fraction
cannot be excluded.  For instance, reacceleration of
\ee \ by shock waves of SNIa may considerably increase $\tau_{SD}$.
Reacceleration seems ``natural'' (charged particles should be accelerated
each time they encounter a shock wave) and is taken into account in some models
of CR propagation (e.g. Strong and Moskalenko 1998, Ptuskin 2001). 
Note that in the case of  standard CR (protons of 
energies $\sim$1 GeV) there is
not much room for reacceleration, just because of energetic arguments. Indeed,
the kinetic 
power of Galactic CR is $\sim$10$^{41}$ erg/s, i.e. a sizeable fraction
of the kinetic energy released by Galactic supernovae ($\sim$10$^{42}$ erg/s).
However, the power in $\sim$10$^{43}$ $e^+$/s (of 1 MeV each) is
$P_{e^+}\sim$10$^{37}$ erg/s. SNIa constitute about 20\% of the total
Galactic SN and collectively release a kinetic power a thousand times larger than 
$P_{e^+}$; thus, 1 MeV positrons may be reaccelerated many times
and their slow-down time  $\tau_{SD}$ largely increased.

On the other hand, the positron confinement time in the disk may be
shorter than the standard value of $\tau_{CONF}\sim$10$^7$ yr.
The reason is that, because of their low energy (and correspondingly low
gyroradius in the Galactic MF) 1 MeV positrons may diffuse very little on the 
density fluctuations of the MF, and thus they may escape more easily than 
the higher energy particles of standard Galactic cosmic rays. 
Indeed, in the standard theory of resonant diffusion of charged particles 
on MF irregularities the condition for diffusion is that the particle 
gyroradius must be larger than the smallest scale of the fluctuations 
(and smaller than the largest scale). If it is smaller than the smallest 
scale, then diffusion is largely suppressed. 

The gyroradius of 1 MeV positrons is $R\sim$ 10$^9$ cm for a field of a few
$\mu$G, like the one in the solar neighborhood. On the other hand, in the
 case of the local interstellar plasma, density
fluctuations have been measured through
pulsar scintillation techniques. Armstrong et al. (1995) find a firm upper
 limit of 10$^{10}$ cm to the smallest scale, but they also stress that 
their measurements are compatible with even lower values. 
Thus, it appears that, within current uncertainties, the gyroradius of 
1 MeV positrons may be smaller than the smallest scale of plasma 
fluctuations. Positrons may then diffuse hardly at all and leak easier 
from the confinement zone, i.e. their $\tau_{CONF}$ may be considerably
 reduced. 
Note that the measurements of Armstrong et al. (1995) concern mainly the
 local plasma, and the situation may be different away from the plane of 
the disk.

Those qualititive arguments are supported by a recent study of 
low energy positron
propagation in the ISM by Jean et al. (2005). Their study
concerns specifically the ISM of the Galactic bulge, but their results  
should also apply to the low density ISM away from the disk, which is mostly 
filled with hot ionized gas.
The authors find that in such a medium positrons of $\sim$1 MeV can travel 
distances up to 5.5 kpc before they annihilate (their Fig. 6 and Table 4). 
Most of that distance ($>$90\%) is covered while positrons propagate in the 
collisional 
regime, i.e. when they do not scatter in the irregularities of the MF. 
If positrons 
can cover such large distances, they can certainly escape from the "confinement zone"
 and reach the bulge, following the lines of the poloidal Galactic MF.

Quantitatively, the SPI measurements,  combined with the theoretical SNIa 
disk and bulge emissivity (Sec. 2) require that $\sim$50\% 
of the positrons escape from the disk and annihilate in the bulge (see Sec. 4
for a reassessment of SPI/INTEGRAL data implications).
If the remaining 50\% annihilates in the disk, 
the resulting Bulge/Disk emissivity ratio
is $\sim$1.2, lower than the value of $B/D=5^{+4}_{-2}$ obtained
by SPI/INTEGRAL analysis in Kn\"odlseder et al. (2005). However, as we show 
in Sec. 4, this is only because relatively thin disk configurations have been
considered in the analysis of Kn\"odlseder et al. (2005), whereas it is physically
reasonable to assume that positrons annihilate in considerably larger volumes.
An extended disk of low surface brightness can hardly be seen in current SPI/INTEGRAL
data;  Kn\"odlseder et al. (2005) recognise that for the case of an extended halo and the
argument holds similarly for a disk. On the other hand,
the analysis of Jean et al. (2005) suggests that MeV positrons travel several
kpc before anihilation, if the hot ISM fills a large fraction of the 
propagation volume (which is the case  away from the disk).
In fact, if a fraction of positrons escapes indeed the disk, as assumed here, it 
would be inconsistent to assume that the remaining  
annihilate close to their sources.  

We conclude then that, in order to explain the SPI data, our model
needs (Eq. 1) $\tau_{SD}$/$\tau_{CONF}\sim$1 
instead of the initially considered  value of 0.10, while
in Sec. 4 we argue that 
the SPI/INTEGRAL data are consistent even with $B/D$ values as low as 0.5.
In both cases, a large increase of  $\tau_{SD}$/$\tau_{CONF}$ above its
``nominal'' value is required. Such an increase could 
be obtained, for instance, by increasing  $\tau_{SD}$ by  a factor of
3 and simultaneously decreasing $\tau_{CONF}$ by a factor of 3.
It is hard to evaluate  whether such large modifications
of $\tau_{SD}$  and   $\tau_{CONF}$  are realistic or not. However we feel
that, the
favorable energetics for positron reacceleration
and the large pathlength of positrons in hot ISM (Jean et al. 2005)
indicate that this possibility is not unrealistic.

\subsection{Can positrons enter the galactic bulge ?} 

If the configuration of the Galactic MF is indeed as assumed in Sec. 3.1, 
positrons that escape the cosmic ray halo are directed towards the
bulge, spiralling and drifting along the lines of the dominant poloidal 
MF, with negligible energy losses. Whether they actually reach the bulge or not
depends on the importance of the ``magnetic mirror'' effect they
undergo in the strong gradient of the dipole MF. That effect is minimised
if, when positrons enter the poloidal field,  their velocity component
 $v_{||}$ 
parallel to the field lines  
is much larger than the corresponding  perpendicular component $v_{\bot}$. 
Otherwise, most of them should be  deflected backwards before reaching 
the bulge (as happens with electrons of the solar wind, entering 
perpendicularly the lines of Earth's magnetic field, which are
 trapped in the van Allen radiation belts).

It turns out that the condition $v_{||} >> v_{\bot}$ may be naturally 
achieved. When positrons are still in the cosmic ray halo, they diffuse on the
 turbulent component of the MF at small scales, but at large scales
 their diffusive motion follows the regular (toroidal) component. That kind of
 motion has been studied by Casse, Lemoine and Pelletier (2002), who found
 that the components of the diffusion coefficient $D$
parallel ($D_{||}$) and perpendicular ($D_{\bot}$) to the regular magnetic
 field $B_0$ are related by
\begin{equation}
\frac{D_{\bot}}{D_{||}} \ = \ \left[ \frac{B^2}{B_0^2+B^2} \right]^{2.3}
\end{equation}
where $B$ represents the mean intensity of the inhomogeneous (turbulent)
 component. This implies that, even close to the disk (where $B\sim B_0$)
 $D_{\bot}\sim$ 0.2 $D_{||}$, whereas near the border of the diffusion zone
 (where  $B < B_0$) one has $D_{\bot} << D_{||}$. In other terms, 
positrons diffuse essentially along the regular component of the MF.
In consequence, their flux $J = -D \nabla n_{CR}$ 
(where $n_{CR} $ is the cosmic
 ray density) is dominated by a component parallel to the lines of
the regular magnetic field. 

The configuration of the Galactic MF can only be continuous between the regions
 where the various components of the regular field dominate. The toroidal
 field changes smoothly into a poloidal one and positrons leaving the
 former enter the latter with a velocity essentially parallel to its 
field lines (since the component $v_{||}$ dominates their   motion).
 For that reason, the magnetic mirror effect should be negligible and 
most positrons escaping the disk
should find their way into the Galactic bulge.

\subsection{Positron annihilation in the bulge}

The amount of gas in the bulge and its properties (density, 
temperature,
ionisation stage etc.) are very poorly known at present and it is hard
to predict how the \ee \ annihilation will take place, although
the observed 511 keV line spectra give some hints to that
(e.g.  Guessoum et al. 2004, 2005; Churazov et al. 2005, Jean et al. 2005). 
Assuming that the bulge is $\sim$10 Gyr old
and has a mass of $\sim$10$^{10}$ \ms, one finds that the mass
return rate from old stars (red giants and AGB stars) is $\sim$0.1
\ms/yr. That gas is expelled at relatively low velocities
(a few 100 km/s), but it is hard to know
the current gas density profile in the bulge, because 
the gas is dissipative and should slowly sink towards a gas torus in the
galactic plane (from which new stars would occasionally form).

A lower limit to the gas mass in the bulge is obtained through analysis  
of infrared data (from IRAS and COBE/DIRBE) concerning the inner, or Nuclear,
bulge by Launhardt et al. (2002). These authors find that $\sim$2 10$^7$ \ms \
of hydrogen reside in the Nuclear Bulge, out to a distance
of $\sim$200 pc from the Galactic center (mostly in an outer torus), 
while another $\sim$4 10$^7$ \ms reside outside that region; in both cases
the gas is mostly cold, dense, and clumpy in nature. Accounting for He and
metals, the authors estimate the total gas mass in the central kpc 
(Central Molecular Zone) to 10$^8$ \ms. The presence of that gas may be 
explained by the action of the Galactic bar, driving gas from the galactic
disk to the inner regions until the gas settles on stable orbits 
(e.g. Binney et al. 1991). However, part of it certainly originates
from the gas slowly released by aging stars of the Bulge.

Assuming that the total gas mass in the bulge is indeed of the order of 
10$^8$ \ms and fills a volume with a radius of 1-2 kpc, the mean gas density
in the bulge should be $n\sim$0.1-1 cm$^{-3}$.  
The average slow-down time scale of 
positrons in the bulge  and the distance they travel before annihilation
depend on the nature of the ISM and the volume filling factors of the various
phases. Since the physical conditions of the bulge ISM are very poorly known at
present, one has to turn the problem around and use the observed spectral
signature of the \ee-e$^-$ annihilation radiation to derive those conditions.
This has been done in Jean et al. (2005) who find that the emission
results from positrons annihilating
in about equal amounts in the warm (T$\sim$8 000 K) neutral and  ionized
phases of the ISM. Their analysis also suggest
that the fraction of annihilation emission
from molecular and hot gas has to be less than 8\% and 0.5\%, respectively.

Those results imply that, either

i) the bulge ISM is dominated by the warm neutral and ionized phases, with
the cold and hot phases having very small filling factors, or

ii) the bulge hot ISM dominates (as is the case in the solar neighborhood)
but its morphology is such that positrons escape it and enter the warm
and denser phases, where they annihilate, or

iii) the bulge hot ISM dominates and, since positrons cannot 
slow down and annihilate
there, they mostly escape outside the bulge; only a small fraction 
of them enter
the warm phase of the bulge and annihilate. The total rate of positrons going 
through the bulge is then much larger than indicated by the detected signal.

Obviously, case (iii), although physically plausible, exasperates the 
difficulty
of finding  a prolific positron source, able to provide a lot more 
than 1.5 10$^{43}$ \ee/s.

Case (ii) is the one favoured in  Jean et al. (2005). However, it is
based on our understanding of the local ISM, while conditions in the bulge
may be different. For instance, the density of 
heating sources of the bulge ISM (SNIa) is smaller than the
corresponding one in the local disk (which is dominated by core
collapse supernovae) and the volume filling factor of 
the hot ISM in the bulge may be much smaller than in our solar neighborhood.
 Case (i) may then be closer to reality.

In any case, the large magnetic field in the bulge (perhaps, 
up to 1mG, according to
Morris and Serabyn 1996, and references therein) indicates an efficient
confinement of positrons there, much more efficient than in the local disk.
Thus, it is expected that positrons entering the bulge will be
trapped and annihilate inside it (unless extremely special conditions, like
e.g. a hot ISM with very large volume filling factor and/or peculiar
configurations of the magnetic field, allow a large fraction of them to escape).

For illustration purposes we assume then that positrons in the
bulge are annihilated in a gaseous medium
which has a density profile similar to the stellar one, assumed here to be a 
simple exponential (i.e. no triaxiality is taken into account)
with a characteristic scalelength R=0.32 kpc.

\section{Disk surface brightness and Bulge/Disk ratio}

In this section we discuss the disk emissivity profile from positron annihilation 
and the resulting Bulge/Disk emissivity and flux ratio.
We assume that the Milky Way emissivity results from a bulge with emissivity 
L=1.2 10$^{43}$ \ee/s (resulting from transfer of $\sim$50\% of the disk SNIa 
positrons plus those produced by the bulge SNIa population) 
and from a disk with emissivity and morphology that are constrained 
from SPI measurements.
For illustration purposes we adopt four different models for the disk:

{\it Model A}: The disk has an  exponential profile, with scalelength of 
2.6 kpc and scaleheight of 0.2 kpc. This is one of the two disk models  
adopted in the SPI data analysis of Kn\"odlseder et al. (2005, model D1). 
It corresponds to an old (but not very old) disk and assumes that positrons 
annihilate close to their SNIa sources. The disk positron emissivity is 
L$_{A}$=0.4 10$^{43}$ \ee/s, i.e.
slightly less than half the remaining disk positrons (after the 
transfer of 50\% to the bulge) annihilate in the disk near their sources, 
while the other half may escape completely from the disk; this model
corresponds to the average disk 
emissivity allowed by the analysis of Kn\"odlseder et al. (2005) for that morphology. 
It is used here as a check of our own modelisation against the work of 
Kn\"odlseder et al. (2005). The Bulge/Disk ratio of that model is 3.

{\it Model B}: The disk has an exponential scalelength of 4 kpc and a
 scaleheight of 1 kpc. Those values reflect  the assumption  that positrons 
diffuse and annihilate away from their sources. In that case, their 
distribution at the moment of annihilation corresponds more to the 
distribution of cosmic rays. In fact, the scalelength of 4 kpc may be even
too low, in view of 
the well known fact that the cosmic ray source distribution 
in the Milky Way has a surprisingly flat profile as a function of 
galactocentric radius (e.g. Strong and Moskalenko 1998). The positron 
emissivity of the disk is L=0.95 10$^{43}$ \ee/s, i.e. it corresponds 
to the remaining $\sim$50\% of the disk \ee \ emissivity from SNIa.
The Bulge/Disk emissivity ratio of this model is $\sim$1.3

{\it Model C}: This model involves two disks (beyond the bulge, which is common
to all models):
The first (Disk-B) is the same as in Model B; the second (Disk-C)  
has a short scaleheight of 0.1
 kpc, a rather large scalelength of 5 kpc and a \ee \ emissivity of 
L=0.3 10$^{43}$ \ee/s. It illustrates the possible contribution from 
the decay of radioactive $^{26}$Al. This isotope is produced mainly 
in massive stars, which are distributed essentially as the gaseous layer. 
The adopted \ee \ emissivity corresponds to the amount of decaying $^{26}$Al 
in the Galaxy, estimated to 3 \ms/Myr (e.g. Diehl et al. 2005).  
This configuration is the one of model D0 (young disk) adopted in 
the analysis of Kn\"odlseder et al. (2005). It assumes that positrons
from  $^{26}$Al decay do not travel away from their sources, since
they propagate in a dense medium (an assumption which can be only
partially justified, since $^{26}$Al producing
supernova heat the ISM; the emitted positrons encounter a hot ionized gas, at 
least in some directions, where they can propagate to large distances).
The Bulge/Disk emissivity ratio of this model is $\sim$1.

{\it Model D}: The disk has larger scalength (6 kpc),
scaleheight (3 kpc), and  positron emissivity
 (L=2.4 10$^{43}$ \ee/s) than Model B.
It is used to show that Bulge/Disk ratios as low as 0.5 
may be comptible with current SPI/INTEGRAL data, provided the disk
is diluted enough.

Only Model A has a bulge/disk \ee \ emissivity ratio compatible with the values obtained 
in SPI/INTEGRAL data analysis by Kn\"odlseder et al. (2005), while the other ones are smaller. 
In all cases, the disk is truncated at an inner radius of 1.3 kpc (since it cannot physically 
co-exist with the bulge, see also Robin et al. 2004) and at an outer radius of 15 kpc (since
 cosmic ray acceleration sources do not exist at such large distances, see also Strong and
 Moskalenko 1998).

At this point one might argue that, even if positrons occupy a large volume,
the 511 keV emissivity should be proportional to the product of their density
times the electron (gas) density; this would result in a thin disk emitting
511 keV photons, not an extended one. However, positrons have a finite 
lifetime (the slow-down time $t_{SD}$, which is the sum
of the lifetimes  in diffusive and collisional 
regime) and in that respect they could be assimilated to radioactive particles:
during that period they can travel away from their sources (in the hot ISM)
and fill a large volume, but once their lifetime 
has ``expired'', they annihilate locally. In that case, {\it 
the resulting 511 keV
profile reflects the distribution of positrons} and not the product of their
density times the gas density.

\begin{figure}
\centering
\includegraphics[width=0.5\textwidth]{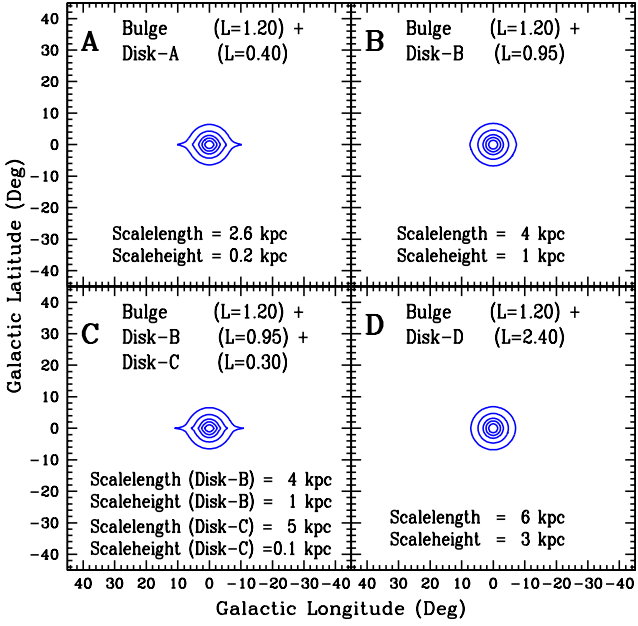}
\caption{\label{f4} Simulated profile of the 511 keV emission 
of the Milky Way in galactic coordinates. In all panels, L represents
positron emissivity in units of 
10$^{43}$ \ee/s and the last
isoflux contour represents 16.7\% of the peak central flux. In each panel, letters
(A, B, C, D) indicate models described in Sec. 4. All those models
have the same bulge emissivity (L=1.2), assumed to result from the transfer
of 50\% of positrons from disk SNIa (plus the ones produced in the bulge,
see text)  and they differ only by the
disk properties (emissivity, scalength and scaleheight) as indicated
in each panel. In all panels, bulge positrons are annihilated locally
in the bulge. In panel A, disk positrons are also annihilated locally, i.e.
near their sources (SNIa, distributed as the old stellar population); 
this model, with
a disk positron emissivity  L=0.4, corresponds to the  disk
found in SPI data analysis (its morphology corresponds to D1 
in Kn\"odlseder et al. 2005).
In panel B, positrons occupy a ``Cosmic ray disk'' of large scalelength
and scaleheight (see text), with a positron emissivity
 L=0.95, i.e. half of the
disk SNIa positrons; this disk would be invisible with currently available 
SPI exposure.
In panel C, a second disk, of small scaleheight (similar to disk D0 of
 Kn\"odlseder et al. 2005)
is added to the configuration of panel B; such a disk would correspond 
to positrons
emitted by Galactic $^{26}$Al (L=0.3) and locally annihilated,  and it 
is marginally 
detectable in current SPI data.
Finally, Panel D is as Panel B, but the disk is more luminous (positron
emissivity L=2.4,  i.e. Bulge/Disk=0.5)
and more diffuse than in case B; 
despite its luminosity, it is also ``invisible'' with
current SPI sensitivity.} 
\end{figure}

\begin{figure}
\centering
\includegraphics[width=0.5\textwidth]{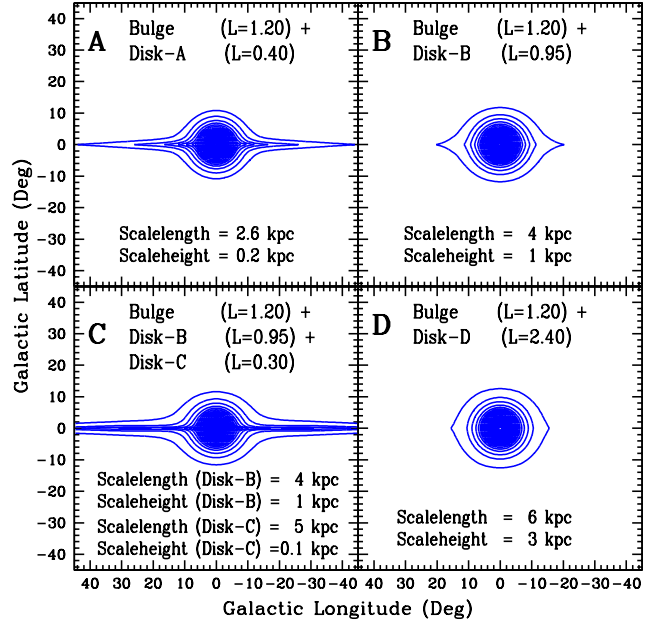}
\caption{\label{f5} Same as Fig. 4, but the outmost isocontour corresponds to
4\% of the maximal flux, i.e. sensitivity is improved by a factor of 4 w.r.t. Fig. 4.
The disks are clearly detected in cases A and C, but only marginally detectable
in cases B and D.}
\end{figure}

The 511 keV flux profile of the Galaxy, in
longitude $l$ and latitude $b$, is calculated by integrating
the volume emissivity $\rho(r,l,b)$ 
of the various model components (bulge plus disk) along the line of sight
\begin{equation}
dF(l,b) \ = \ \int_0^\infty \rho(r,l,b) \ dr \ dsinb \ dl
\end{equation}
where $r$ is the distance from the Sun (e.g. Prantzos and Diehl 1996).
Bins of 0.5 deg are used for both longitude and latitude. 
The volume photon emissivity $\rho_{\gamma}(r,l,b)$ (in photons cm$^{-3}$ 
s$^{-1}$)
is related to the  positron emissivity $\rho_{e^+}(r,l,b)$ (in \ee \ 
cm$^{-3}$  s$^{-1}$) by
\begin{equation}
\rho_{\gamma} \ = \ [2(1-f)+\frac{1}{4} \ 2 \ f] \ \rho_{e^+}
\end{equation}
where $f$ is the positronium fraction. This expression translates 
the fact that
positrons may annihilate either directly, with a probability (1-$f$),
giving 2 photons
of 0.511 MeV, or after positronium formation with probability $f$. 
Positronium is formed 1/4 of
the time in the singlet $^1S_0$  state (which gives again 2 photons of 
0.511 MeV)
and 3/4 of the time in the triplet  $^3S_1$ state (which gives 3 photons
with energies covering the range 0-0.511 MeV).

The gamma-ray flux is calculated by assuming that the positronium 
fraction is $f$=0.93, as in the analysis of Kn\"odlseder et al.
 (2005); note that the recent spectroscopic analysis of the SPI data 
by Jean et al. (2005) suggests $f$=0.967$\pm$0.022, which changes 
the resulting fluxes by 10\%: for $f$=0.93, Eq. (4)
leads to $\rho_{\gamma}$=0.6 $\rho_{e^+}$, where for  $f$=0.967 one obtains
$\rho_{\gamma}$=0.54 $\rho_{e^+}$.

In all cases, the total positron emissivity of a galactic component (bulge
or disk) occupying a volume $V$ is normalised to the assumed value of L, i.e.
\begin{equation}
L \ = \ \int_V \ \rho_{e^+} \ dV
\end{equation}

The results of our simulations appear in Fig. 4, where the properties
 of each model
also appear in the corresponding panel. Isocontours are drawn at equal levels
(in linear scale) corresponding to 1/6 of the peak central
flux  (i.e. the outer contour is at 16.7\% of the peak flux).
Assuming that the peak flux in the central 3 square degrees is 
$\sim$4 10$^{-4}$ cm$^{-2}$ s$^{-1}$ (see Fig. 7), the outmost
contour corresponds to a flux of $\sim$7 10$^{-5}$ cm$^{-2}$ s$^{-1}$.
This corresponds approximately to the current sensitivity
of SPI/INTEGRAL for an extended source at 511 keV (Kn\"odlseder et al.
2005).

It can be seen that in case A (already tested in Kn\"odlseder et al.
2005), the disk is marginally detectable, whereas in Cases B and D
the disk is ``invisible'' with current SPI sensitivity. Case C is
also interesting, since both a thin and a thick (and more luminous) disk
can be compatible with current data.

\begin{figure}
\centering
\includegraphics[width=0.5\textwidth]{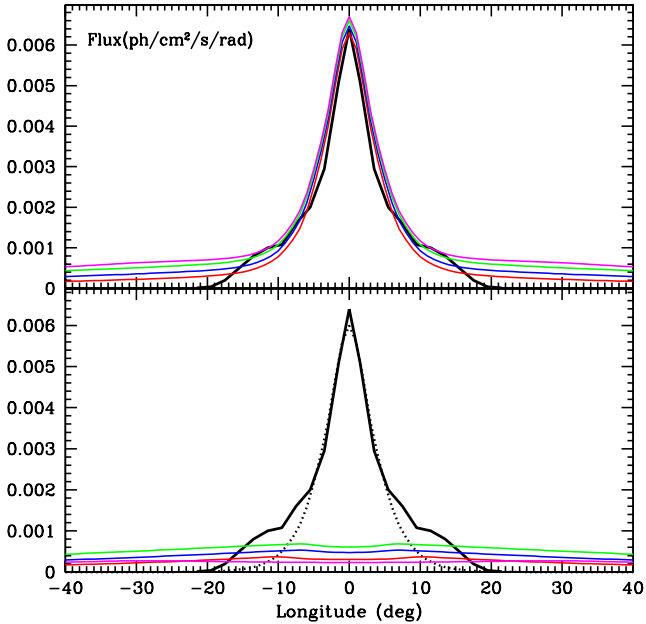}
\caption{\label{f4} Flux in the 511 keV line as a function of Galactic 
longitude,
corresponding to the configurations of Figs. 4 and 5. In both panels, 
the thick solid curve is the
one derived from SPI data analysis (Kn\"odlseder et al. 2005, Fig. 5); 
it is centered to l=0$^o$
and assumed symmetric w.r.t. the Galactic center. In the upper panel, 
the combined
flux of each model is given, 
while in the lower panel, the flux from each individual 
component is given;
the {\it dotted curve} in the lower panel represents the bulge component,
 common in all models.
In order to reproduce exactly the results of Kn\"odlseder et al. 2005 
(their Fig. 5),
fluxes are integrated in latitude, for $|b|<$30$^o$.} 
\end{figure}

Fig. 5 shows the effect of increasing sensitivity (by a factor of 4) 
for the same model configurations. The outmost contour is
at 4\% of the central flux and now the disks are clearly seen
in cases A and C. There are hints for a disk in Case B and
indications for a flattened bulge in case D. Clearly, deeper observations of 
the Milky Way are required to understand the various aspects of positron
propagation in the Milky Way.

Fig. 6 displays the corresponding fluxes as a function of Galactic longitude.
They are integrated in latitude, for $|b|<$30$^o$, to be directly
comparable to the data displayed in Fig. 5 of Kn\"odlseder et al.
 (2005). In the upper panel the combined fluxes are given
(i.e. those directly corresponding to Figs. 4 and 5), while
in the lower panel, the fluxes of individual components are displayed;
the latter allows, in particular to appreciate the contribution of the bulge
component, always dominating the inner $\pm$12$^o$ of longitude.
All models give flux profiles compatible with the SPI data. The differences
concern mostly the region outside $|l|\sim$20$^o$, where the extended and more
luminous models C and D display  larger fluxes than models A and B; still
the differences are well within the uncertainties of current SPI data.

Finally, Fig. 7 displays integrated fluxes from square regions of the 
sky centered
on the  Galactic center, i.e. for $(|l|, |b|)<$Angle $A$, as a function of $A$
($|b|$ runs from 0$^o$ to 90$^o$ and 
 $|l|$ runs from 0$^o$ to 180$^o$).
Differences between the various models appear more clearly in that diagram.
They are negligible for $(|l|, |b|)<$10$^o$, of the order of 15\%
for $(|l|, |b|)<$20$^o$ and they reach a factor of 2 when integration in
the full $(|l|, |b|)$ range is performed (i.e. over 4 $\pi$). The latter
case, at Angle=180$^o$, is compared to the range of models fitted to
SPI/INTEGRAL data by Kn\"odlseder et al. (2005, 
vertical error bars on the right).
Models A (already tested in Kn\"odlseder et al. 2005) and B (with a bulge/disk
ratio of 1.2 only)
are well within the current measurement uncertainties.
Models C and D  have slightly larger total fluxes (by 10\%) but only because
such extended disks have not been used in the SPI data analysis in 
Kn\"odlseder et al. (2005).

\begin{figure}
\centering
\includegraphics[angle=-90,width=0.45\textwidth]{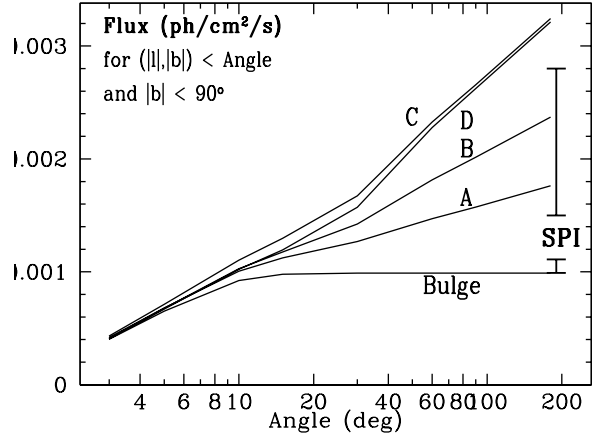}
\caption{\label{f7} Flux in the 511 keV line inside longitude $l$ and
latitude $b$, such as : $|l|$, $|b|$ $<$Angle, as a function of Angle
($|b|$ runs from 0$^o$ to 90$^o$ and 
 $|l|$ runs from 0$^o$ to 180$^o$).
Curves indicate results of our modelling, for the
adopted bulge and models A, B, C and D, respectively. The results from
the SPI data analysis (Kn\"odlseder et al. 2005) are displayed for the full range
of their composite models ({\it upper vertical line})
and for the corresponding bulge component only ({\it lower vertical line}), 
integrated over 4 $\pi$.}
\end{figure}

\section{Summary}

In this work, we investigate several aspects of the positron annihilation 
line observed in the Milky Way, and we propose a model that satisfies the
current observational constraints.

In Sec. 2 we reassess the positron production rate from SNIa, on the basis
of  recent data. We find that (a) the combined bulge+disk production is slightly 
larger than required by observations, but (b) the bulge/disk ratio is the inverse 
of what is observed, 
{\it if positrons are assumed to annihilate close to their sources}.
Point (a) may be just a coincidence (i.e. SNIa may not be the Galactic
 positron sources), 
but it is argued here that there is a way to reconcile points (a) and (b).

It is suggested that about half of the disk positrons may escape the disk 
and be transported
 via the Galactic magnetic field to the bulge, where they annihilate.
In Sec. 3.1, the (still poorly known) configuration of the Galactic MF is 
presented briefly;
 the existence of a strong poloidal component, suggested by recent data, 
is crucial to the 
success of the model. In Sec. 3.2, some aspects of the propagation of
 positrons in the hot
 and low density ISM far from the disk are considered; in such conditions
 positrons (even 
of MeV energies, such as those produced by radioactivity) may travel for
 several kpc before 
slowing down and annihilating. They may then escape the disk and be 
chanelled to the bulge 
by the poloidal MF of the Galaxy, which dominates away from the disk. 
In Sec. 3.3 it is 
argued that positrons may enter the bulge avoiding the mirror effect 
(due to the MF gradient),
 since their motion is always dominated by a velocity component parallel
 to the MF lines.
In Sec. 3.4 the annihilation of positrons inside the bulge (where the 
configurations of the
 MF and of the ISM are also poorly known) is discussed.

In Sec. 4 we calculate  sky maps of the 511 keV emission, based on 
various assumptions about
 the extent of the positron annihilation region. It is argued that, in general (and in the
 framework of our model, in particular) positrons have to annihilate away from their sources.
 We show quantitatively that the SPI/INTEGRAL data are fully compatible even with bulge/disk positro emissivity
 ratios lower than 1, provided that sufficiently (but not unreasonably) extended positron 
distributions are considered. We stress, in that respect, that positrons
can be assimilated to radioactive particles (due to their finite
slow-down time), so that {\it 
the resulting 511 keV
profile reflects the distribution of positrons} and not the product of their
density times the gas density.

Thus, SNIa may indeed be the dominant positron source in the Milky Way, 
as thought for many 
years. The rates, positron yields and galactic distribution of other candidate sources
(e.g. X-ray binaries, millisecond pulsars, microquasars etc.)
are much more poorly known than those of SNIa. However, if SNIa turn out to produce much 
less positrons than claimed in Sec. 2, the arguments of Sec. 4 may be used for any other
 positron sources, which have sufficiently large yields but not the correct bulge/disk
 ratio. Indeed, if the positron yields of some of those sources are large enough, the 
fraction required to be transferred to the bulge may be small (and even zero, i.e. the
 positrons of the disk have just to move sufficiently far away and even escape the 
Galaxy, in order for their annihilation to be undetectable). 

The model proposed here relies heavily on our poor understanding of the Galactic
magnetic field and of the propagation of low energy positrons in it. However, its 
assumptions may  be tested,
through future observational and theoretical developments.
Systematic multi-wavelength studies of SNIa, including the infrared, will determine
 ultimately the typical positron yield of those objects. A small
511 keV emission outside the bulge is currently seen by SPI/INTEGRAL
(Kn\"odlseder, private communication) and, given enough exposure, the spatial extent 
of that emission will be determined (either by INTEGRAL or by a future instrument);
 an extended disk emission will prove that positrons travel indeed far away from their
 sources. Finally, the morphology of the Galactic magnetic field, and especially the 
presence of a poloidal component, will be put on more sound basis through further
 measurements (e.g. Han 2004).

\acknowledgements{ I am grateful to Martin Lemoine, Pierre Jean and Jurgen Kn\"odlseder 
for illuminating discussions, and to an anonymous referee for a very constructive report.}

\def\aj{AJ}
\def\apj{ApJ}
\def\apjs{ApJS}
\def\aap{A\&A}
\def\aaps{A\&AS}


{}

\end{document}